# Carbon cluster diagnostics-III: Direct Recoil Spectroscopy (DRS) *versus ExB* Velocity Analysis

Shoaib Ahmad[1,2,*], S. A. Janjua[1], M. Ahmad[1], B. Ahmad[1], S. D. Khan[1], Rahila Khalid[1]

[1]*Accelerator Laboratory, PINSTECH, P.O. Nilore, Islamabad, Pakistan*

[2]*National Centre for Physics, Quaid-i-Azam University Campus, Shahdara Valley, Islamabad, 44000, Pakistan*

Email: sahmad.ncp@gmail.com

## Abstract

A comparative study is being reported on the identification of heavy carbon clusters using two different techniques namely the Direct Recoil Spectroscopy (DRS) and velocity spectrometry using a Wien ExB velocity filter. In both the techniques carbon clusters are formed in situ either under energetic heavy ion bombardment of graphite surfaces for which DRS is employed or these are emitted from the regenerative sooting discharges and analyzed using ExB velocity analysis.

## 1. Introduction

The purpose of the present study is to investigate the relative merits of two different mass diagnostic techniques namely the Direct Recoil Spectroscopy (DRS) and velocity analysis by using an $E \times B$ filter. The two diagnostic techniques have been employed on the identification of carbon clusters that range from $C_2$ to moderately large ones $C_m$ (m ≥ 100). At PINSTECH various carbon cluster production techniques have been developed with appropriate mass diagnostics. The first technique that was developed involved production of soot containing a wide range of carbon clusters $C_m$ (2 ≤ m ≤ 100) on the graphite surface by heavy ion bombardment at oblique angles. This sooted surface was diagnosed by energy analyzing the direct recoils emanating from binary collisions. The direct recoil spectra of carbon clusters provided not only the information about the constitution of the sooted surface; it also gave important clues regarding the mechanisms responsible for the production of these clusters [1]. The other important aspect of DRS in our experiments has been the simultaneous use of the heavy ion beam as the agent of soot formation and as projectiles for the direct recoils emanating from the surface.

The other series of experiments involved the creation of regenerative sooting discharges with graphite hollow cathodes in noble gas environments. Carbon clusters are formed in the carbonaceous vapour that is established in the glow discharges at low pressures and high voltages. A recent review discusses the mechanisms and processes of the formation of the regenerative soot [2]. A compact, permanent magnet based $E \times B$ velocity filter was designed and used for the detection and identification of carbon clusters extracted from the discharge. The velocity spectra delivered by $E \times B$ filter is tuneable and can have higher resolution under certain circumstances. When combined, these two techniques can offer unique advantage in mass analyses of heavier species.

In this comparative study, we present the relative diagnostic abilities of the two above mentioned methods. PINSTECH's 250 keV heavy ion accelerator was used for these studies. Whereas, the energy of our accelerator is appropriate for inducing an optimum sputter-deposited sooted surface, it is on the lower side to be used as a DRS probe on its own for a wider mass range of targets. On the other hand, the accelerating voltage is sufficient to have high enough resolution for the detection of very large clusters $C_m$ ($2 \leq m \leq 1{,}000$).

## 2. Direct Recoil Spectroscopy (DRS)

In DRS the energy analysis of the recoiling ions is performed in binary encounters with the incident ions at large recoil angles. A home-made electrostatic energy analyzer was used for energy analysis. The bombarding ions initiate cluster forming mechanisms and can also eject the existing ones by imparting them sufficient energy as direct recoils. The recoil spectra are composed of clusters with greater than or less than the mass of the incident ion and have a strong dependence on the emission or recoil angle as well as the ion energy. The constituents of a sooted surface recoil in binary collisions as DRs which carry characteristic energy $E_{DR}$ as a function of the target to projectile mass ratio $m_2/m_1$, angle of the direct recoil $\theta_{DR}$ and projectile's energy $E_0$. The energy of clusters of mass $m_2$ recoiling at angle $\theta_{DR}$ is given by [3]

$$E_{DR} = 4\frac{m_1 m_2}{(m_1 + m_2)} 2 E_0 \cos^2 \theta_{DR}.$$

These DRs being the primary events of the energetic projectile – target interactions subsequently initiate collision cascades in which the energy is shared with other nearest neighbours. Most of the sputtered species result from the interaction of these collision cascades with the surfaces. The direct recoils have well defined energies, trajectories and points of origin whereas, the sputtered species have broad energy distribution peaked at half the binding energy of the surface atoms ~ $E_b/2$. Both the mechanisms are involved in the ion-induced cluster formation on the surface, however, the energy spectra of DRs can be easily identified against the background of low energy sputtered species. Furthermore, the charged DRs have been easily discriminated against the predominantly neutral sputtered clusters.

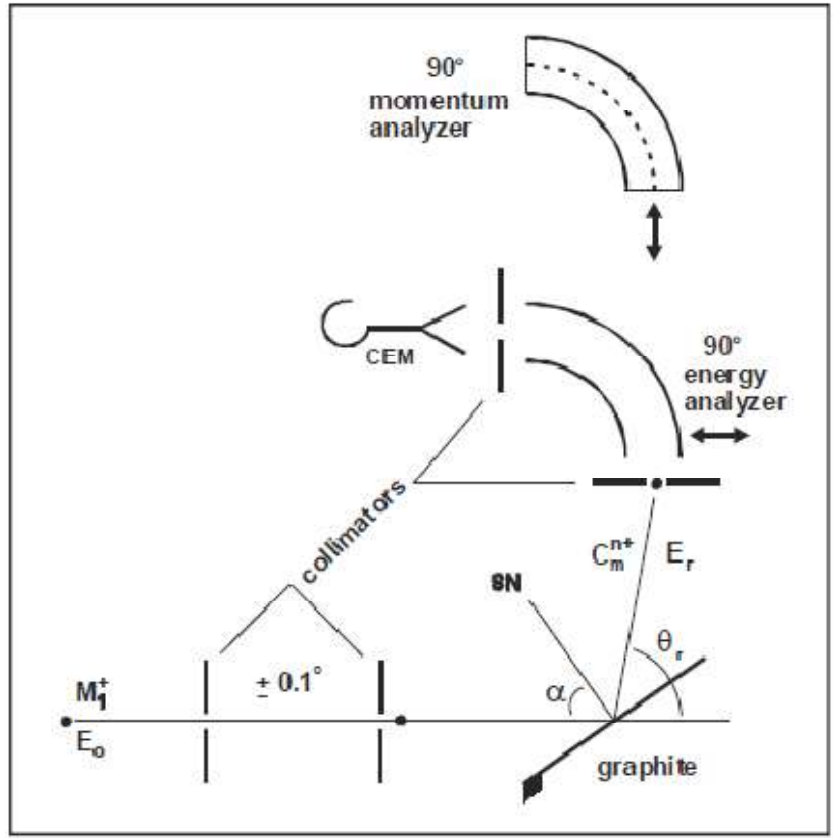

**FIG. 1. Schemetic diagram of the Direct Recoil Spectrometer.**

## 2.1. Experimental Setup

Figure 1 shows the experimental arrangement for the Direct Recoil spectrometer for the production as well detection of carbon clusters under energetic heavy ion bombardment with mass $M_1^+$ and energy $E_0$. The projectiles make angle α with the surface normal SN. Large values of α imply higher sputtering yields and shallower depths for the implanted ions. DRs with energy $E_r$ are allowed to enter the energy analyser at large recoil angle $\theta_r$ (≥ 80º). A channel electron multiplier (CEM) is used to detect the positively charged clusters $C_m^{n+}$, where n indicates charge multiplicity. A set of collimators is used to restrict the beam divergence to less than 0.1°.

## 2.2. Clusters Recoiling from Heavy Ion Bombarded Graphite Surface

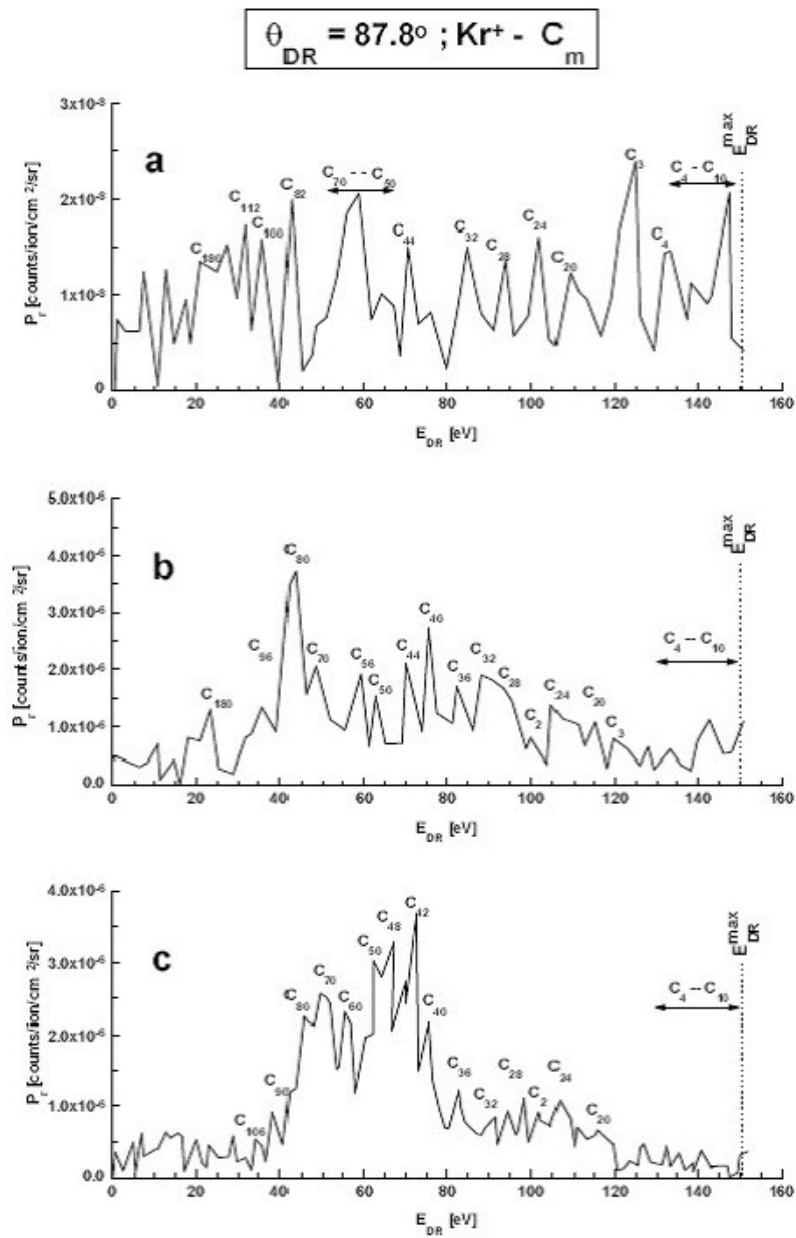

**Figure 2. Direct Recoil spectra emitted from the 100 keV irradiated Kr$^+$**

FIG. 2 shows emission of a wide range of clusters from Kr$^+$ irradiated graphite surface as a function of the ion dose. The clusters recoil with characteristic energies $E_{DR}$. FIG. 2a,b and c present the dynamic behaviour of the formation and fragmentation of various clusters. In the figure, $E_{DR}^{max}$ refers to the maximum transferable energy to a cluster in direct recoil calculated from the above equation. The three consecutive spectra were taken with 4 μA beam incident at a grazing incidence angle ~ 80º. Each of the three spectra is taken after consecutive ion dose ~$10^{14}$ ions. The gradual build up of the higher clusters have been identified as fullerenes in which non-abutting pentagons dominate the structures.

The ion induced clustering phenomenon at the graphite surfaces can be attributed to the physical and chemical changes that occur when heavy ions are stopped via nuclear and electronic collisions. We have shown elsewhere [4] that the ratio of nuclear to electronic stopping $S_n/S_e$ has a direct bearing on the energetics of C-C bond formation and fragmentation processes. The generation and detection of carbon clusters with the technique of Direct Recoil Spectroscopy under energetic heavy ion bombardment provides an addition to the already existing techniques for the study of clusters. This can be easily extended to other materials and useful information obtained due to the unique feature of simultaneous production and detection of clusters on the target surface by the same ion beam. Therefore, one can study the ion dose effects as well as the processes and mechanisms of formation and fragmentation of all sorts of clusters in DRS.

## 3. *ExB* Velocity Spectrometry

A compact, permanent magnet based $E \times B$ velocity filter was designed and used for the detection and identification of carbon clusters extracted from the discharge. The main emphasis during the design and fabrication of the velocity filter was to keep the overall size to within manageable dimensions and to be of low cost. Neodymium iron boron bar magnets of 14x14 mm cross section and 45 mm length were used in the construction of the permanent magnet that provides 180 mm constant field region. The velocity spectra delivered by $E \times B$ filter is tuneable and can have higher resolution under certain circumstances

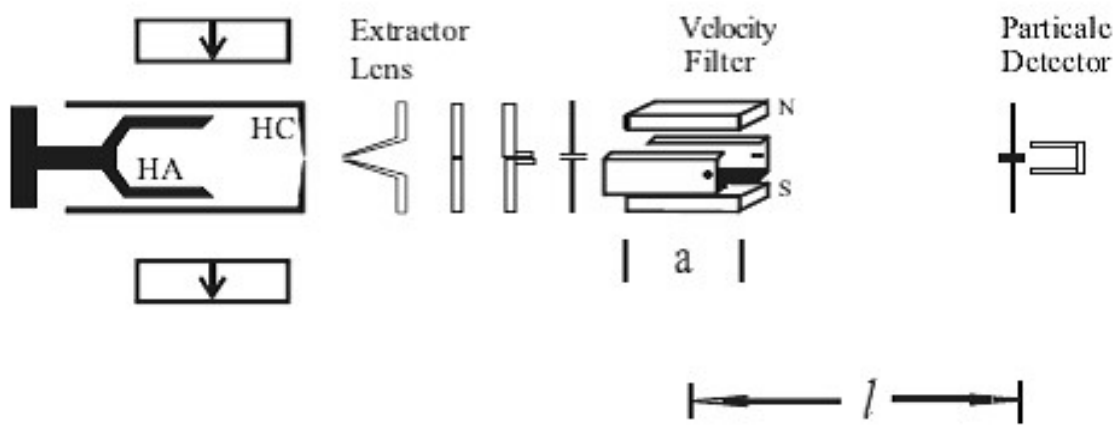

**Figure 3. The Velocity Filter as heavy carbon cluster mass spectrometer.**

Figure 3 shows the cluster ion source, extraction lens and the velocity spectrometer of dimensions *a*. The Faraday cage is at a distance *l* from the $E \times B$ filter. The cluster ion source is a specially designed carbon cluster ion source [5] composed of a hollow cathode shown as HC, hollow anode (HA) and a set of six neodymium iron boron (NdFeB) bar magnets indicated with arrows in the figure. This permanent magnet based $E \times B$ velocity filter performs mass analysis in a characteristic way by deflecting all clusters according to their respective masses. The compensating electric and magnetic fields can allow masses with velocities $v_0 = B_0 / \varepsilon_0$ where $B_0$ is the magnetic field (= 0.35 T) at the axis and $\varepsilon_0$ is the compensating electric field which is the only variable in this arrangement of constant magnetic field velocity filter.

Dedicated $E \times B$ velocity filters are designed with specific mass ranges. Since higher compensating electric fields $\varepsilon_0$ are required for lighter particles for a given energy and $B_0$, the limit is set on the one hand by the lightest mass to be detected with the highest velocity $v_{max}$ and the largest mass with the minimum velocity $v_{min}$ on the other. The requirement for the detection of heavy carbon clusters imposes the acquisition of highest $B_0$ that sets all other parameters accordingly. The resolution is directly proportional to the product of $al$. Thus the resolution can be enhanced for a given spectrometer by increasing only $l$ that is the length of the free flight region.

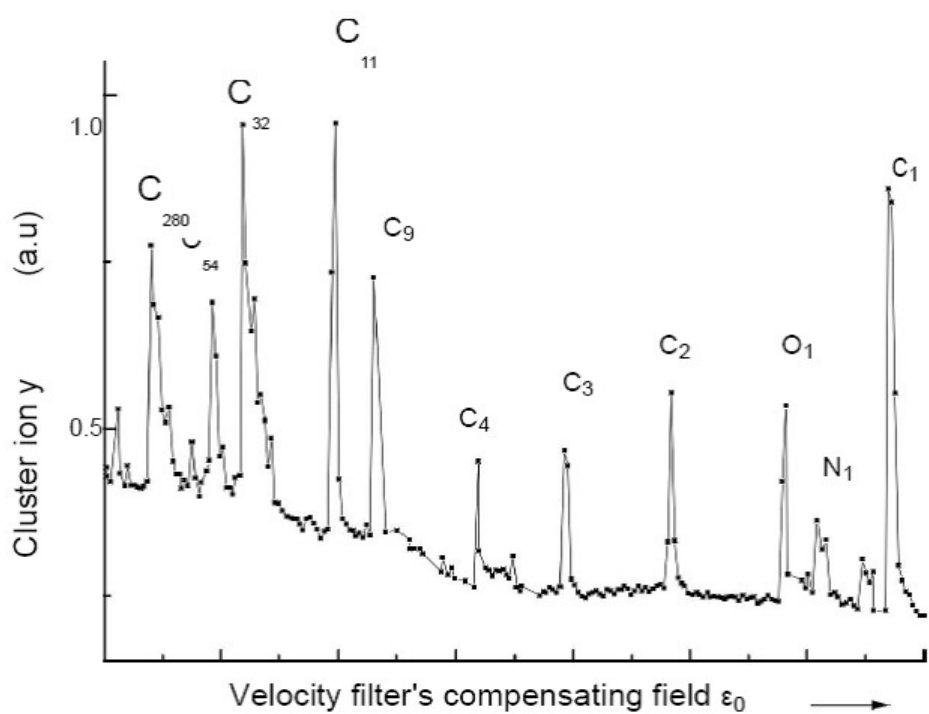

**Figure 4. The velocity spectra of carbon clusters emitted from a well sooted source.**

## 3.2. Velocity Spectra of Carbon Clusters emitted from Regenerative Soot

FIG. 4 shows the velocity spectrum of clusters that are emitted from a well sooted hollow graphite source. The extraction voltage is 1 kV and such a small value was chosen as it is needed to spread out the smaller clusters. The larger ones are grouped together at the lower compensating voltages. The resolution is poor for the larger clusters while for the smaller ones it is sufficient to identify individual clusters. The resolution of the velocity analyzer is variable within the same spectrum for different masses. It also depend on the extraction voltage and to resolve the larger clusters one has to go to higher energies but at the expense of loss of information about the smaller ones. One has to strike a balance between various requirements and then optimize the experimental parameters.

The resolution of the $E \times B$ velocity filter is determined by the dispersion d of masses $m_0 \pm \delta m_0$ from the resolved mass $m_0$ that travels straight through the filter with velocity $v_0$ (= $B_0/\varepsilon_0$). Dispersion $d \propto al\,(\delta m_0/m_0)(\varepsilon_0/V_{ext})$ where $a$ and $l$ are the lengths of the velocity filter unit and the flight path, respectively. For a given ratio $\delta m_0/m_0$, the dispersion $d$ can be enhanced by stacking multiple filter units since $d \propto na$, $n$ being the number of filter units or increasing the flight path $l$ and also by enhancing $V_{ext}$.

## 4. Relative Comparison of the two Diagnostic Techniques

A comparison of the two widely different spectroscopic techniques i.e., DRS and velocity analysis for the identification of the ion-induced clusters formed on the graphite surface and in the regenerative soot are presented in Table 1.

TABLE I: Comparison of DRS and Velocity Analysis

| DRS | Velocity Analysis |
|---|---|
| DRS provides recoiling clusters' energy spectrum from which masses are deduced. | Cluster mass is directly available from the velocity spectra. |
| The mass identification depends on the resolution of the energy spectrometer. | Resolution can be improved by increasing the extraction energy or the drift length. |
| The evolution of clusters on the bombarded surface can be monitored as a function of time and ion dose. | The cluster evolution when monitored continually provides the parametric dependence of the regenerative soot. |
| The technique is limited to the heavy ion bombarded targets and has a limited utility as a general mass diagnostic tool. | It is a general technique and can be employed to masses varying from low to very heavy ones. |

Our comparative study highlights the relative merits of the two techniques for the identification of the mechanisms of cluster formation. In the case of ion-induced sooting of the surface of the graphite target, DRS serves as a powerful cluster diagnostic tool. While for the regenerative soot formed in graphite hollow cathodes, velocity spectrometry is a well deserved analysis method. Although these had earlier been used separately and independently for mass analysis on various accelerator driven systems in many labs, we have found these twin techniques to complement each other when used in the experimental setups described in this presentation.